# Enhanced Tunnel Spin Injection into Graphene using Chemical Vapor Deposited Hexagonal Boron Nitride

*M. Venkata Kamalakar\*, André Dankert, Johan Bergsten, Tommy Ive, Saroj P. Dash†*

Department of Microtechnology and Nanoscience, Chalmers University of Technology, SE-41296, Göteborg, Sweden
\*venkata.mutta@chalmers.se; †saroj.dash@chalmers.se

*The van der Waals heterostructures of two-dimensional (2D) atomic crystals constitute a new paradigm in nanoscience. Hybrid devices of graphene with insulating 2D hexagonal boron nitride (h-BN) have emerged as promising nanoelectronic architectures through demonstrations of ultrahigh electron motilities and charge-based tunnel transistors. Here, we expand the functional horizon of such 2D materials demonstrating the quantum tunneling of spin-polarized electrons through atomic planes of CVD grown h-BN. We report excellent tunneling behavior of h-BN layers together with tunnel spin injection and transport in graphene using ferromagnet/h-BN contacts. Employing h-BN tunnel contacts, we observe enhancements in both spin signal amplitude and lifetime by an order of magnitude. We demonstrate spin transport and precession over micrometer-scale distances with spin lifetime up to 0.46 nanosecond. Our results and complementary magnetoresistance calculations illustrate that CVD h-BN tunnel barrier provides a reliable, reproducible and alternative approach to address the conductivity mismatch problem for spin injection into graphene.*

**Keywords:** Hexagonal boron nitride, Graphene, Spintronics, Tunnel spin injection, CVD, Spin transport, Spin precession, Tunnel barrier, 2D atomic crystal, Conductivity mismatch

## I. INTRODUCTION

The discovery of graphene[1,2] and several other 2D materials[3] has opened up the possibility of employing their fascinating multifunctional properties for nanoelectronic and spintronic applications.[4–7] Graphene in particular is an ideal material for spintronics[8] due its predicted high electron spin coherence time and diffusion length.[9,10] Since the first report on spin transport and precession in graphene[11], vigorous studies are being pursued to achieve these theoretical goals.[12,13] Electrical spin injection into graphene from a ferromagnetic (FM) metal is achieved by employing an insulating tunnel barrier, which circumvents the intrinsic conductivity mismatch problem.[14] Inspired by the famous tunnel magnetoresistance structures, ferromagnetic tunnel contacts with metal-oxide barriers are often used for spin injection and detection in graphene.[15,16] The conventional metal-oxide barriers such as $Al_2O_3$ and MgO grown on graphene suffer from pin-holes, interface roughness and defects related to oxygen vacancies and doping in graphene.[17–19] Even though these barriers offer a certain level of interface resistance, there has been a lack of reproducible evidence for tunneling transport. Although studies involving $TiO_2$ seeded MgO barriers[17] showed tunneling behavior, such hybrid metal-oxide barriers still suffer from roughness, non-crystallinity and reproducibility issues.[20] These factors greatly influence spin polarization and spin lifetimes in graphene.[17–19] Therefore, the quest is still on for a better spin tunnel barrier, a prime building block in determining the future course of graphene spintronics.[21]

Here, we explore the role of h-BN as a tunnel barrier for spin injection into graphene. In recent years, there has been a growing interest in van der Waals heterostructures of 2D crystals such as graphene with hexagonal boron nitride (h-BN). In fact, van der Waals heterostructures can lead to the observation of many interesting phenomena such as tunnel transistors, metal-insulator transition, itinerant magnetism, Wigner crystallization.[22–24] Hexagonal boron nitride is an insulating isomorph of graphene with a bandgap of 6 eV [25] and a lattice mismatch of only 1.7 %.[26] Being a 2D crystal, it is ideally devoid of dangling bonds, atomically smooth, and possesses high mechanical strength and thermal conductivity, as well as even higher chemical stability than graphene. The h-BN has been successfully used as a substrate to improve mobility[27,28] and to achieve long distance spin transport in graphene.[29] Also, it has recently been employed as a tunnel barrier in graphene-based transistors.[22–24] The latter experiments revealed the possibility to precisely control the resistance by varying the number of atomic layers. The 2D structure of the h-BN makes it appealing for spin-tunnel barriers because of the absence of surface states, which rules out any charge or spin traps at the interface.[30] Such 2D layers can also act as diffusion barriers, which protect the underlying graphene from possible doping during the deposition of ferromagnetic contacts and subsequent processing.

Recent theoretical studies by Kapran et al. showed that the tunnel contacts with tailored RA and large spin polarization values can be achieved by engineering the number of h-BN layers at the junction.[31] Studies also predict tunnel magnetoresistance (TMR) exceeding 100 % and perfect spin filtering with h-BN tunnel barriers.[32] In a recent experimental attempt, Yamaguchi et al.[33] used exfoliated single layer h-BN for spin injection into graphene with spin signal of a few milliohms (with spin polarization < 2 %) and spin lifetime ~ 50 ps. The low resistive h-BN barrier used in their experiments lead to no considerable improvements when compared to ferromagnet/graphene transparent contacts.[34] Also, the exfoliated h-BN used in these studies are smaller in size, which complicate the fabrication scheme and is unsuitable for wafer-scale processing.[22–24,28] Therefore, in order to establish the significance of h-BN tunnel barriers for spin injection, it is necessary to demonstrate possibilities of achieving higher spin signals and spin lifetimes in graphene. At the same time it is also important to explore the viability of CVD grown h-BN as tunnel barrier for both charge and spin based devices, which has prospect for wafer scale fabrication. We address both these challenges in our work by employing CVD grown h-BN as a tunnel barrier for spin injection into graphene.

In this letter, we show reliable and reproducible tunneling behaviour of large area CVD grown h-BN and demonstrate efficient spin injection, transport and precession in graphene. Furthermore, we report an enhancement of the spin signal and spin lifetime by using resistive h-BN tunnel barriers, which circumvents the conductivity mismatch problem. Our observations supported by calculations elucidate that h-BN tunnel barrier resistances can be tailored to achieve desirable magnetoresistance values. While the present work focuses on the graphene spintronic devices, our results are generic for a large class of h-BN charge and spin based tunnel devices.

## II. RESULTS

We illustrate the underlying concept and spin injection device structure in Fig. 1a and 1b, which integrates the heterostructures of 2D materials (Graphene and h-BN) with ferromagnetic contacts. The spin-polarized current is injected through a Co/h-BN ferromagnetic tunnel junction, creating a non-equilibrium spin accumulation and splitting of the chemical potential ($\Delta\mu$) in graphene (Fig. 1a). The resulting spin polarization is detected nonlocally by another h-BN/Co tunnel contact placed at a distance of typical spin diffusion length (~ 2 μm). Using such van der Waals heterostructures, we demonstrate tunneling characteristics of large area CVD h-BN, tunnel spin-injection signal at room temperature with a micrometer-length spin transport and Hanle spin precession in graphene.

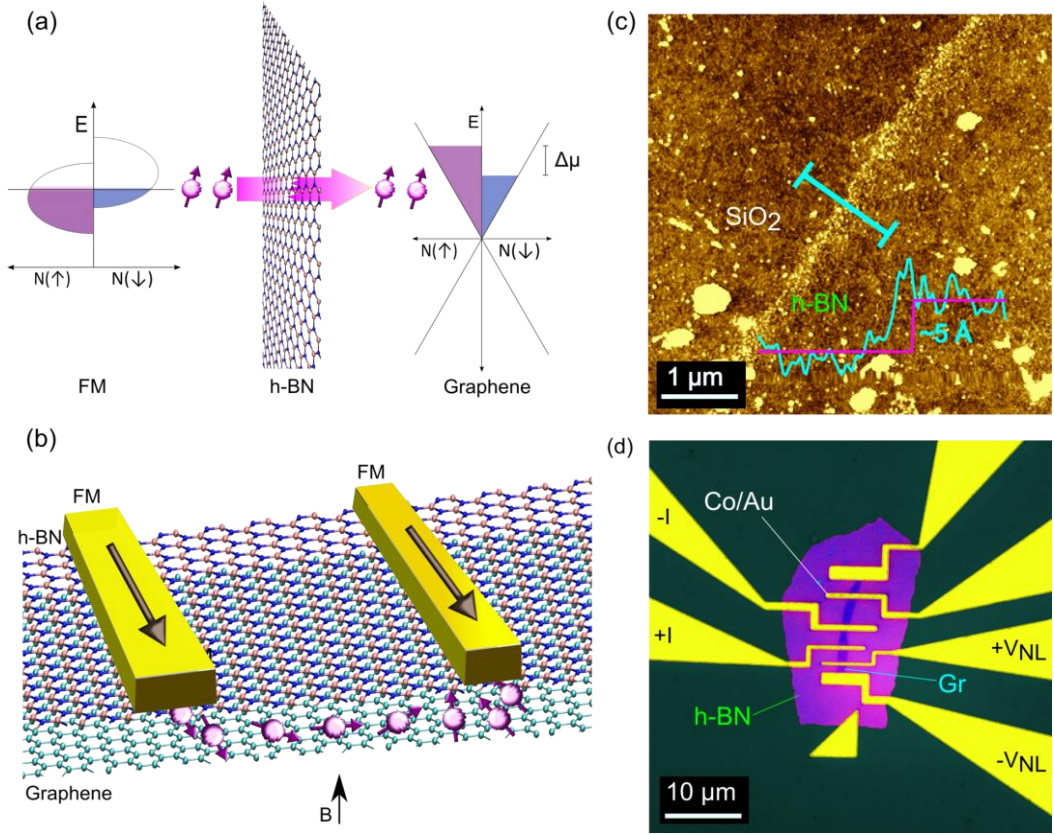

**Figure 1| Hexagonal boron nitride (h-BN) tunnel barrier for graphene spintronics**. **a,** Scheme showing spin injection into graphene through h-BN tunnel barrier. **b,** Schematic representation of the injection of spin-polarized electrons from a ferromagnet into graphene through an atomically thin h-BN/ferromagnet tunnel structure. **c,** Atomic Force Microscopy image of a CVD h-BN layer transferred onto a Si/SiO$_2$ substrate. **d**, Colored optical microscope image of a multiterminal spin-transport device showing a heterostructure of a graphene (Gr.) flake covered by a h-BN tunnel barrier and ferromagnetic cobalt electrodes patterned by electron-beam lithography.

**Tunneling characterization of hexagonal boron nitride -** We fabricated graphene spin-transport devices using h-BN as tunnel barrier and ferromagnetic Co electrodes for spin injection and detection. We prepared graphene flakes on Si/SiO$_2$ substrate by exfoliation from HOPG using scotch tape method[1]. Subsequently, we transferred the CVD grown layer of h-BN using a wet transfer process (elaborated in Supplementary information). We have been able to achieve a ripple free transfer of CVD h-BN over large areas on our Si/SiO$_2$ chips. The h-BN was then patterned to regions well covering the graphene flakes (see Supplementary information Fig. S1 and Fig. S2). The position of the Raman shift for the CVD h-BN has been reported to be ~ 1369 cm$^{-1}$ by the supplier[35], which matches well with the thickness based studies on exfoliated h-BN.[36] We also performed AFM (atomic force microscopy) characterization revealing a minimum thickness of 5 Å for the h-BN layer as shown in Fig 1b, which is the effective thickness observed for single layer h-BN.[22,36,37] In our AFM based analysis, we also found the thickness of the h-BN to be about ~ 5-10 Å (Supplementary information Fig. S3) with few regions having even higher thickness. It has been reported that the resistance area product (RA) of h-BN scales exponentially with the number of layers.[22] Thus, the RA scaling analysis is an efficient method to resolve the thickness of h-BN, especially when the number of layers lies between 1-3. In our experiments, we used AFM and resistance scaling analysis to ascertain the atomically thin nature of the CVD h-BN. Because of the variation in mean thickness, we also found the RA of the Co/h-BN contacts to vary between devices. Nevertheless, we found that the contacts in a particular device have reasonably similar RA indicating a uniform local thickness of h-BN layer.

We fabricated devices with and without h-BN tunnel barriers. In the following description we refer h-BN0 as the device without any tunnel barrier, and h-BN1, h-BN2, h-BN3 and h-BN4 as devices with h-BN tunnel barriers in the order of increasing RA. The details of the device fabrication are described in the supplementary information. A representative optical micrograph (color coded with enhanced contrast) of a device with a graphene flake covered by a patterned atomically thin layer of CVD h-BN and several ferromagnetic electrodes fabricated by electron-beam lithography is shown in Fig. 1d.

We performed the electrical characterization of the devices in different measurement geometries to ascertain the tunneling behavior of the Co/h-BN/graphene contacts. The I-V characteristics measured in two-terminal (2T) and three-terminal (3T) configurations are presented in Fig. 2a. The channel characterization is done by regular 4-probe measurements and the spin signal is measured in nonlocal (NL) geometry. Fig. 2b shows the 2T and 3T tunneling I–V characteristic curves of a h-BN tunnel contact with RA ~ 10 kΩμm$^2$. The IV characteristics fit well (the lines in Fig. 2b) to Brinkman-Dynes-Rowell (BDR) model[38] for an asymmetric barrier. The fittings provide effective barrier heights of φ ~ 1.57 eV for 2T and 1.49 eV for 3T. These values match with the barrier heights for h-BN from previous reports[22,39] and are close to typical barrier heights for oxide tunnel barriers used in magnetic tunnel junctions (MTJs). The weak temperature dependence of the tunnel resistance is a robust criterion for a tunnel junction without pin holes and trapped states[40]. Such weak temperature dependence for the h-BN tunnel contact at zero bias and 170 mV is displayed in Fig. 2c. In addition, we also tested metal/h-BN/metal junctions where we obtained tunneling IV characteristic curves fitting well to the BDR model apart from their weak temperature dependence. We find a variation in the RA product from device to device, which can be primarily attributed to the variation in mean thickness of CVD h-BN (5 -10 Å) as observed in AFM measurements. Fig. 2d shows the different RA obtained in our devices with h-BN tunnel barrier (Co/h-BN/Graphene) and without h-BN (Co/Graphene) tunnel barrier. We note that tunneling characteristic curves with weak temperature dependence have also been seen in our other devices (h-BN1, h-BN2, h-BN3 and h-BN4, for example I-V curves for h-BN2 is shown in supplementary information Fig. S4 and Fig. S5). These facts point out that our h-BN tunnel barrier is of good quality, and devoid of pinholes and interface states. The RA of the h-BN tunnel contacts in our devices is found to be between 2 -13 k$\Omega$ μm$^2$, where we measured the spin signal. These values correspond to a thickness in the range of 1-2 layers of h-BN in light of the recent thickness dependent resistance studies[22] (Supplementary information Fig. S6). The values of the RA for the Co/h-BN contacts are orders of magnitude larger when compared to transparent Co contacts in device h-BN0 (~ 400 $\Omega$ μm$^2$). In spite of the challenges associated with the growth and transfer of CVD h-BN, we have been able to observe the tunneling behavior of most of the contacts over a large scale, in a wide range of temperature. The graphene channel resistivities in our devices are found to be in the range of 2–6 k$\Omega$/□ with linear I–V behavior. We observed characteristic gate dependent Dirac curves for the graphene channels of our devices with a mobility of 2000-3000 cm$^2$ V$^{-1}$ s$^{-1}$. The details of the graphene characterization are presented in Supplement information (Fig. S7).

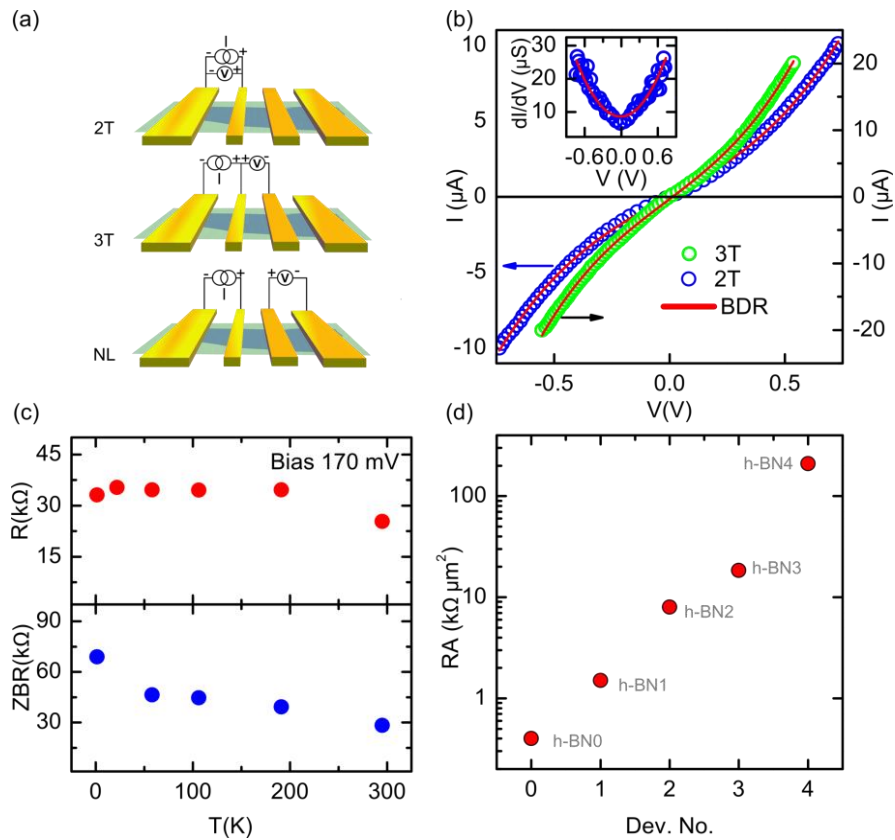

**Figure 2| Spintronic device and electrical characterization**. **a**, Circuit configurations for 2-terminal (2T), 3-terminal (3T) and nonlocal (NL) measurement **b**, 2-terminal and 3-terminal IV characteristics of a Co/h-BN/graphene contact with fits to Brinkman Dynes and Rowell (BDR) model. The inset shows the BDR fit to the dI/dV of 2-terminal measurements **c**,

Temperature-dependence of h-BN tunnel resistance measured in 3-terminal geometry at 170 mV (Top) and zero bias (ZBR) (Bottom). **d,** Resistance area product (RA) (at zero bias) of the injection tunnel contacts obtained on different devices.

**Spin transport and precession in Graphene with h-BN/Co tunnel contacts -** Having established the excellent tunneling behavior of the Co/h-BN contacts, we focus on spin-dependent transport measurements carried out on our graphene devices. The measurements were performed in a nonlocal (NL) spin-valve geometry, where the charge current path is isolated from the spin diffusion signal measured by the voltage probes, as shown in the inset of Fig. 3a. In a typical device, we inject current between the +I and –I contacts, and measure the nonlocal voltage signal between the +V and –V contacts. The spins injected through the Co/h-BN contacts accumulate in the graphene channel and diffuse laterally, thereby being detected at the nonlocal voltage probes. The nonlocal resistance is recorded while the in-plane magnetic field is swept from a negative to a positive value, followed by a reverse sweep. Ferromagnetic electrodes having different shape anisotropies (as shown in Fig. 1d) were employed as injectors and detectors to achieve different magnetization switching fields. The measurements were performed with a DC current in the range of ±10 µA.

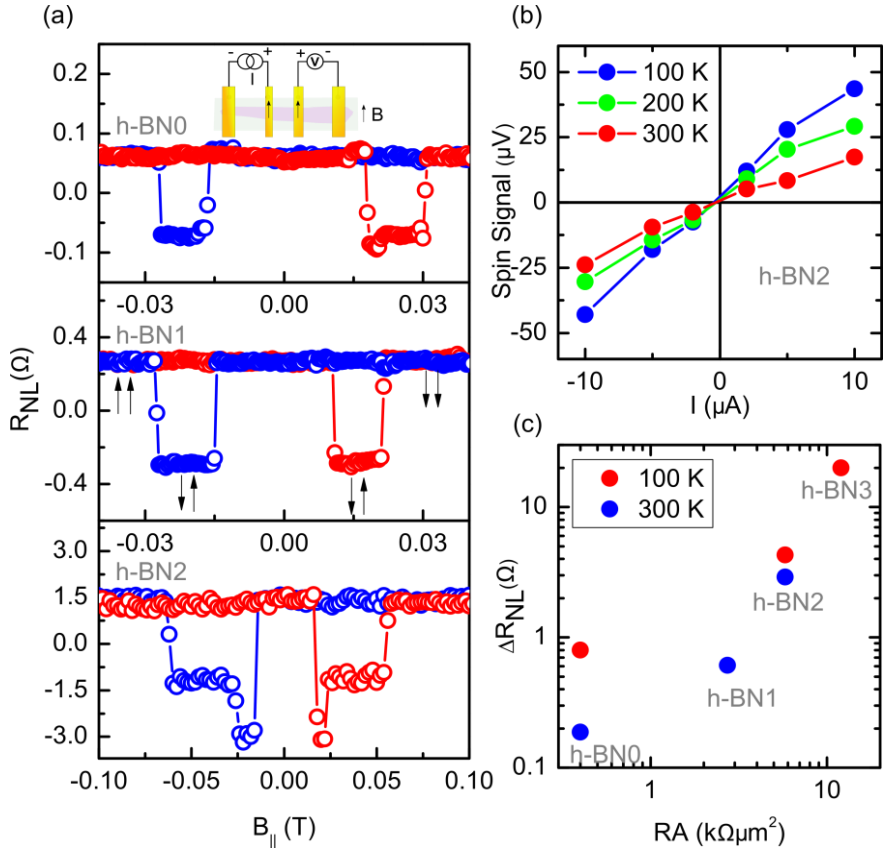

**Figure 3| Spin transport in graphene at room temperature**. **a,** Nonlocal spin-valve signal ($R_{NL} = V_{NL}/I$) measured at room temperature for h-BN0 (device with no tunnel barrier), h-BN1 and h-BN2 (devices with different h-BN tunnel resistances) using an injection current of I = +5 µA (Inset: Nonlocal measurement configuration). Sweep directions of the magnetic field are indicated by blue (positive to negative field sweep) and red (negative to positive field sweep) colors. The magnetic configurations of the injection and detection electrodes are illustrated for both sweep directions in the plot for h-BN1 **b,** Nonlocal spin-valve voltage

signal (ΔV = $V_p$ - $V_{ap}$) as a function of injection bias current for device h-BN2 at temperatures 100 K, 200 K and 300 K. **c**, Enhancement in spin signal ($\Delta R_{NL}$) with increase in tunnel contact resistance (RA) in different devices.

The room temperature nonlocal spin valve measurements for devices with a h-BN tunnel barrier (h-BN1 and h-BN2) and without a tunnel barrier (h-BN0) are shown in Fig. 3a. The zero bias resistance values of the devices are presented in Fig. 2d. We observed nonlocal spin-valve switching for both parallel and antiparallel configurations of the injector and detector electrodes in our devices (with the magnetic configuration shown for h-BN1 in Fig. 3a). The additional switching in device h-BN2 in the low field range is due to spin current reaching the outer reference Co electrodes. Notably, the magnitude of the spin signal is enhanced from h-BN0 to h-BN1, h-BN2 and h-BN3 with increasing contact RA. In all the devices, we observed the spin signal to depend linearly on the injection current. We note that we did not measure a spin signal in our h-BN4 device with large tunnel contact resistance. A representative bias dependence of the spin signal for h-BN2 at different temperatures is shown in Fig. 3b. This confirms that the spin accumulation in the graphene channel created by the tunnel spin injection from the Co/h-BN contact is in the linear response regime. The enhancement in the nonlocal spin-valve resistance $\Delta R_{NL}$ with contact resistance RA for different devices is shown in Fig. 3c. We note that the spin signal shows a monotonic increase in value as the h-BN tunnel contact resistance is increased. The insertion of a high resistive h-BN tunnel barrier not only prevents the backflow of spins into the ferromagnetic contact, but also minimizes the spin flip scattering at the interface leading to a higher spin accumulation in graphene and hence a higher spin signal. We describe later (in the nonlocal spin signal calculation) how the $\Delta R_{NL}$ is quantitatively related to the tunnel barrier resistance.

In order to evaluate the spin lifetime of the electrons in graphene devices with h-BN tunnel barrier, we performed nonlocal Hanle spin precession measurements by sweeping the magnetic field perpendicular to the device geometry while the magnetization axes of the injector and detector electrodes are being kept parallel. The injected spin-polarized electrons precess about the perpendicular magnetic field with Larmor frequency ($\omega_L$) while diffusing towards the nonlocal detector contact. In Fig. 4a, we show the nonlocal Hanle signal measured for devices with h-BN tunnel barrier (h-BN2 and h-BN3) and without tunnel barrier (h-BN0) at different temperatures for a bias current of +5 µA. In addition to the increase in the magnitude of the spin signal we also observe a decrease in line width of the Hanle curves. For device h-BN3 with contact RA = 12.6 kΩ µm², we observed a nonlocal Hanle resistance with $\Delta R_{NL}$ ~ 10 Ω at 100 K. This would correspond to a nonlocal spin valve signal of 20 Ω. To remove possible non-spin related effects, we have symmetrized our raw Hanle data by taking the average of the negative and the positive magnetic field sweeps.[19] To estimate the spin lifetime, we fit the measured data using the Hanle spin transport equation (Eq. 1), involving the spin diffusion, precession, and relaxation factors. The variation of nonlocal resistance ($\Delta R_{NL}$) due to Larmor precession of the spins diffusing towards the detector from the injector is given by

$$\Delta R_{NL} = \pm \frac{P^2 D}{W \sigma_s} \int_0^\infty \frac{1}{\sqrt{4\pi D t}} \, e^{-\frac{L^2}{4Dt}} \, \cos(\omega_L t) \, e^{-(t/\tau_s)} \, dt \qquad (1)$$

where $P$ is the spin polarization of the h-BN/Co contacts (assumed to be the same for both the injector and the detector), and $\omega_L = \frac{g\mu_B}{\hbar} B_\perp$ (with Lande's g-factor g = 2) is the Larmor frequency of the precession of electrons in a perpendicular magnetic field $B_\perp$. $L$ is the effective channel length between the injector and the detector (2 µm in our devices), with a width $W$ and square conductivity $\sigma_s$ of the graphene strip. The diffusing spin-polarized electrons have a lifetime $\tau_s$ and diffusion constant $D$ with a spin diffusion length $\lambda_{sf} = \sqrt{D\tau_s}$.

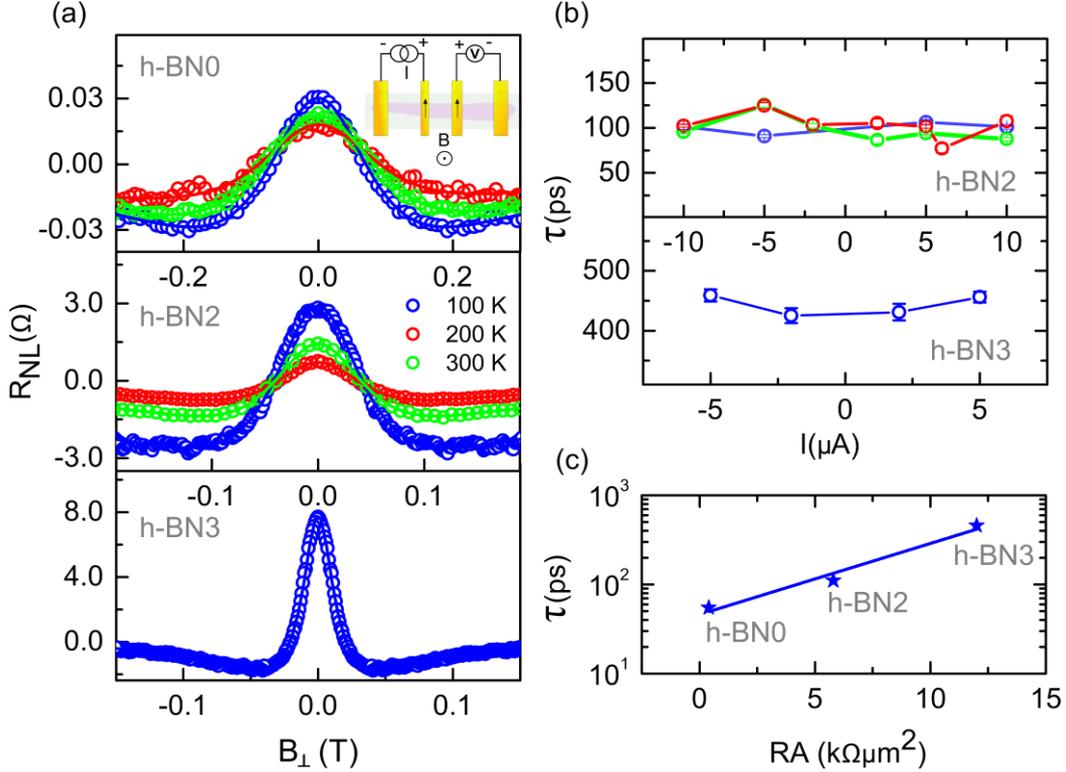

**Figure 4| Spin-precession measurement in graphene**. **a,** Nonlocal Hanle spin-precession signal measured for devices with different tunnel barrier resistances (h-BN0, h-BN2, and h-BN3) as a function of the perpendicular magnetic field, keeping magnetization of the ferromagnetic electrodes in parallel in-plane configuration at 100 K, 200 K and 300 K. **b, Top:** Bias dependence of the spin lifetime for device h-BN2 at different temperatures. **Bottom:** Bias dependence for h-BN3 at 100 K. **c,** Dependence of spin lifetime on the contact resistance extracted from h-BN0, h-BN2 and h-BN3 at 100 K.

The spin lifetime in graphene devices with h-BN tunnel barrier (h-BN2, h-BN3) are found to be much higher than the device without tunnel barrier (h-BN0). When compared with h-BN0 having $\tau_s \sim 50 ps$, the evaluated $\tau_s$ are higher for devices h-BN2 ($\tau_s \sim 110 ps$ with RA ~ 5.8 kΩ µm$^2$) and h-BN3 ($\tau_s \sim 0.46$ ns with RA = 12.6 kΩµm$^2$). The maximum spin signal and spin lifetime were obtained for the h-BN3 device, giving rise to a spin diffusion length $\lambda_{sf}$ of 1.4 µm with a spin polarization P of 14 % and $\tau_s \sim 0.46$ ns. The diffusion constant obtained from the Hanle fitting is D = 0.00416 m$^2$/s. The spin lifetime $\tau_s$ is found to be independent of the bias current for our devices as shown in Fig. 4b. We further observe that for h-BN2, the spin lifetime is also fairly independent of the temperature (top panel of Fig. 4b). In this case the bias current does not influence the diffusive spin transport in graphene much. The weak temperature dependence of the spin lifetime also indicates that the spin relaxation in the channel is mainly dominated by the impurity and charged scattering centers in graphene. Most importantly, the magnitude of the spin signal and the spin lifetime are found to increase with tunnel barrier resistance. A plot showing the enhancement in the spin lifetime with barrier resistance is displayed in Fig. 4c. Such behavior can be understood considering the various sources of scattering. The spin relaxation time ($\tau_s$) is primarily a measure of spin-flip scattering ($\tau_{sf}^{-1}$) inside the graphene channel. In the case of spin injection from a ferromagnet to graphene, the spin escape rate ($\tau_{esc}^{-1}$), a rate by which the spins diffuse back into the ferromagnet and get relaxed, greatly influences the spin lifetime ($\tau_s^{-1} = \tau_{sf}^{-1} + \tau_{esc}^{-1}$). A tunnel barrier overcomes this problem by reducing the backflow probability (ideally, $\tau_{esc} \to \infty$). In addition to this, the spin accumulation at the interface suffers from further relaxation due to spin flip scattering at the ferromagnet-graphene interface.[18] The resistance of the tunnel barrier is a measure of the effective isolation to avoid backflow and spin flip scattering at the interface. Thus an increase in h-BN tunnel barrier resistance results in the reduction of contact induced spin relaxation[17,34], leading to an enhancement in the spin signal and lifetime. However, an indefinite increase of barrier resistance is not viable and we discuss about the upper range of tunnel barrier resistance for maximum spin accumulation in the following.

**Nonlocal and local magnetoresistance calculations -** To understand the effect of barrier resistance on the overall spin signal, we calculate the nonlocal spin signal as a function of the tunnel contact resistance ($R_T$) following the drift-diffusion based analytical formulation by Takahashi and Maekawa.[41] Fig. 5a displays the variation of the spin signal with tunnel resistance for a graphene nonlocal spin valve, calculated using the obtained polarization and spin lifetime of h-BN3. The curve indicates that the nonlocal spin valve signal increases with increasing tunnel barrier resistance and saturates beyond an optimum value of the tunnel resistance ($R_{ij} = \sqrt{R_i R_j}$, $R_i \sim R_j$, $i \to$injector, $j \to$detector). The value of the nonlocal spin valve signal $\Delta R_{NL}$ for h-BN3 (~ twice the Hanle signal in h-BN3) is marked by a red dot along with the theoretical curve. The agreement between the calculated spin signal (which is dependent on tunnel resistance) and observed spin valve signal implies that the values of obtained polarization and spin lifetime are reliable. The plot also reveals that the tunnel resistance $R_{ij}$ in our device h-BN3 lies close to an optimum value for the observation of the maximum spin signal. Beyond the contact induced relaxation regime, any further increase in tunnel barrier resistance will not only saturate the spin signal but also results in higher power dissipation in the device. In addition to the saturation of the spin signal, an enhancement in the tunnel barrier resistance inevitably increases the dwell time ($\tau_n$) of spins inside the channel. This leads to a down turn (as $\tau_n > \tau_s$) in magnetoresistance (MR) as observed in spin valves. In Fig. 5b, we show the 2-terminal MR calculation for a channel with the spin parameters of device h-BN3 using Fert-Jaffrè formula.[14] This calculation shows the possibility of observation of a significant MR in a narrow range of contact resistances with $R_{ij}$ of h-BN3 being closer to maximum. In the calculation, the low MR for small contact resistances is due to the conductivity mismatch, preventing spin-injection from the Co contact into graphene. The decay of the MR for large contact resistances is due to relaxation of the spins during the time spent in the graphene channel (dwell time), a time that becomes very long for large contact resistances. Note that the values of spin parameters such as the spin polarization might also depend on the tunnel barrier resistance ($R_T$) which was not taken into account in these theoretical models. In spite of the fact that this might lead to slight deviation on either side of $R_T \sim R_{ij}$, the constant values that we employed here are quite reliable in the regime where $R_T \sim R_{ij}$ as evident from the match in the non-local calculation. This analysis elucidates that CVD grown h-BN is a reliable solution to the conductivity mismatch problem without requiring the deposition of any oxide based tunneling barriers on graphene. It should be noted that the optimum tunnel resistance ($R_T^{max}$) for the observation of a maximum spin signal/MR also depends on the spin resistance ($R_{ch} \propto \lambda_{sf}$) and thus the quality of the channel. This implies that channels supporting larger spin diffusion length or a higher spin lifetime can overcome the dwell time limit set by the tunnel barrier resistance. With epitaxial graphene channels, Dlubak et al.[13] have shown large spin signal values with a high contact resistance, which has been attributed to the better quality of the epitaxial graphene. Therefore, it might be possible to observe higher MR values with better quality graphene while using h-BN layers as tunnel barriers having higher RA values. Furthermore, with the improvements in the h-BN CVD growth process, it should be possible to achieve precise control over the RA of the tunnel contacts.

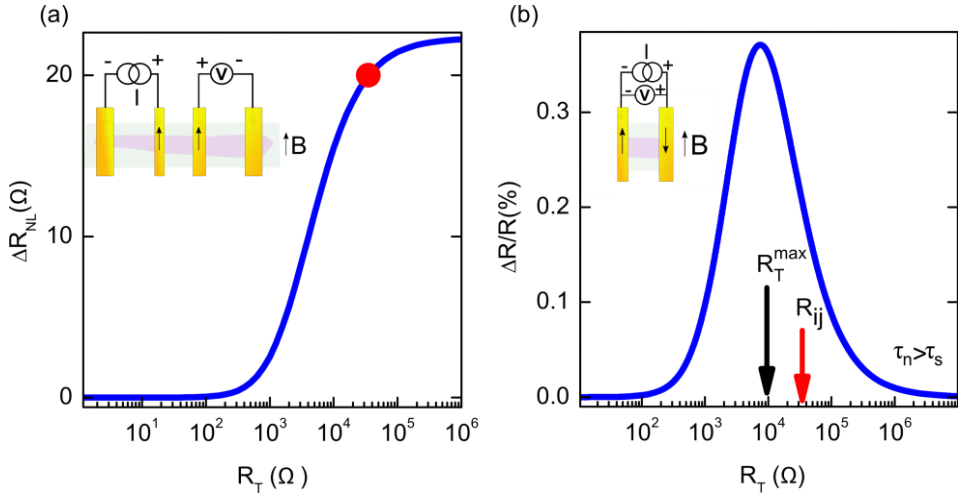

**Figure 5|** **Calculated spin signal as a function of tunnel contact resistance.** **a**, Nonlocal resistance vs. tunnel barrier resistance with the experimental value marked by the red dot. Inset: nonlocal configuration. **b**, Dependence of 2-terminal magnetoresistance (MR) on the tunnel barrier resistance with the arrow in the plot indicating the tunnel resistance ($R_{ij}$) in device h-BN3. Inset: 2 terminal configuration.

## III. DISCUSSION

The current study using CVD h-BN tunnel barrier has several implications. CVD grown 2D crystals hold the key for wafer scale processing and their technological implementation. In our study we observed reliable and reproducible tunneling behaviour of large area CVD h-BN. Often, it has been a challenge to observe performance in CVD grown materials even comparable to their exfoliated counterparts, mostly due to the imperfection introduced during their growth and subsequent transfer processes. In spite of these challenges, we have been able to observe tunneling behavior of CVD h-BN in all devices with h-BN coverage over a large area. On the other hand, most of the metal oxide barriers used in this field do not provide reproducible tunneling nature except for a few reports.[17,42] The tunneling characteristics are reproducible in our CVD h-BN devices despite the observed variation in the contact RA due to the spatial variation in thickness, which we have exploited to investigate the enhancement of spin signals as a function of the tunnel resistance.

The values of spin lifetime of 0.46 ns and polarization of 14 % obtained in our devices show that h-BN is a promising tunnel barrier for spin injection into graphene. This performance is more than an order of magnitude higher than the recent report by Yamaguchi et al. using exfoliated h-BN as tunnel barrier. The lower values of spin lifetime and spin signal in their study can be attributed to the low resistance of the tunnel contact because of the single layer of h-BN used in their experiments. Exploiting the variation of thickness in CVD h-BN and employing resistive h-BN contacts, we successfully demonstrated the enhancement and evolution in both spin signal and spin lifetime in our devices with tunnel resistance. The spin parameters using CVD h-BN are also comparable to that observed using conventional MgO and $Al_2O_3$ barriers.[11,17] The widely observed spin lifetimes vary between 100-500 ps[34], except for occasional reports of spin lifetimes up to 1 ns and above in pristine and hydrogenated graphene.[34,43,44] The fact that CVD h-BN can serve as a tunnel barrier capable of overcoming the conductivity mismatch problem over a large scale together with its potential makes it highly favorable. Due to the 2D nature of h-BN, it is highly compatible with graphene and bypasses the disadvantages of conventional metal-oxide barriers related to growth and doping in graphene.

In addition to the quality of the tunnel barrier, the spin lifetime and polarization depend to a great extent also on the intrinsic scattering mechanisms inside the graphene medium. The observed spin lifetime and diffusion length are significant for our graphene flakes on $SiO_2$ substrate with a carrier mobility of 2000–3000 $cm^2 V^{-1} s^{-1}$, which can be enhanced further by improving the mobility of the graphene.[11,12,29] This, combined with a better quality of CVD h-BN certainly holds the key to answer the important questions in graphene spintronics. The prospect of atomically thin h-BN for spin-tunnel injection goes well beyond this demonstration. Smooth and atomically flat h-BN tunnel barriers are expected to show higher spin-lifetimes and spin-injection efficiencies in graphene than their metal-oxide counterparts. This is highly intriguing, and encourages more detailed spin-transport studies with h-BN tunnel barriers on high mobility graphene devices such as h-BN encapsulated graphene or suspended graphene channels with boron nitride support under the contacts. Recent studies elucidated the use of 2D graphene barrier for spin injection in silicon[45] and magnetic tunnel junctions.[46] However, graphene, being a semi-metal with zero bandgap, is a low resistive barrier. In contrast, h-BN is a high bandgap insulator with a similar structure and can support spin injection into materials with larger spin resistances ($R_{ch}$) such as graphene itself. Opportunities are growing with the proposal by Karpan et al.,[30] showing layers of h-BN can be combined to the perfect spin filtering properties of Co/MLG (multilayer graphene) to realize Co/MLG/h-BN ideal spin injectors. In addition, tunnel contacts with tailored RA and spin polarization values can be achieved by engineering the number of graphene and h-BN layers at the junction. [30] Such heterostructures of graphene and h-BN have a great potential for spintronics in general and are certainly systems for more interesting studies in the future.

## IV. CONCLUSIONS

In conclusion, we have demonstrated tunnel spin injection into graphene using atomically thin CVD hexagonal boron nitride (h-BN) at room temperature. The CVD h-BN showed reproducible tunneling characteristics with weak temperature dependence of the tunnel resistance over a large scale. Taking the advantage of the variation in the thickness of CVD h-BN, we have been able to investigate the evolution of the spin transport with resistance of the h-BN tunnel contacts. With resistive h-BN tunnel contacts, we observed an enhancement in spin transport with spin lifetimes up to 0.46 ns, yielding a spin polarization of 14 %. In light of the magnetoresistance calculations, we show that the CVD h-BN is a reliable and unique way of creating tunnel resistances, which overcomes the conductivity mismatch problem. The integration of CVD h-BN to graphene opens up new avenues to explore the role of h-BN atomic planes for spin injection into other 2D advanced materials,[3,47] metals,[48] semiconductors,[49,50] and magnetic tunnel junctions,[15,16] possibly leading to the observation of large magnetoresistance effects.

**Conflict of Interest:** The authors declare no competing financial interest.

**Acknowledgement:** The authors acknowledge the support from colleagues of Quantum Device Physics Laboratory and Nanofabrication Laboratory at Chalmers University of Technology. The authors would like to thank Jie Sun and Niclas Lindvall for sharing the recipe for transfer process and Samuel Lara Avila and Ram Shankar Patel for useful discussions. This research is financially supported by Nano Area of Advance program at Chalmers University of technology.

**Supplementary Information:** Fabrication details of Graphene/h-BN/Co spin transport devices, Characterization of h-BN tunnel barriers (Electrical measurements, Tunnel barrier quality, Scaling of h-BN tunnel resistance), Gate voltage dependence of graphene and tunnel barrier.